# Novel approach to use the Kelvin Probe method ex-situ for measuring the electron emission yield of insulator materials subjected to electron irradiation

Alexander Marcello Cavalca de Almeida[1,2], Mohamed Belhaj[1], Sarah Dadouch[1], Nicolas Fil[2],

[1] DPHY-ONERA, Toulouse university, Toulouse, France
[2] CNES Toulouse, 18 av. Edouard Belin, 31401 Toulouse, France

E-mail: Alexandre_marcello.cavalca_de_almeida@onera.fr

## Abstract

Measuring the total electron emission yield of dielectric materials remains a challenging task. Indeed, the charge induced by irradiation and electron emission disturbs the measurement. It is therefore important to quantify this charge during the measurement. Using a Kelvin probe allows both the emission yield and the induced charge to be measured. However, this method requires the probe to be placed inside the vacuum chamber, which is often complicated or even impossible. We propose a complete redesign of this method to overcome this issue. A capacitive coupling now allows the potential probe to be placed outside the chamber, in ambient atmosphere. Beyond this major simplification in implementation, we have also introduced several improvements that simplify the measurement protocol and reduce the overall measurement time. The new method was first validated on a metallic sample (Cu), and subsequently applied to a polymer (Kapton).

## 1. Introduction

The study of electron emission under electron impact has been of significant interest and remains a relevant topic of research for several decades [1]. This phenomenon becomes particularly important for materials under irradiation, such as those used in particle accelerators [2], satellites [3], nuclear fusion [4], and particle detectors [5]. Electron emission occurs when a charged particle, here an electron, interacts with the matter; from an energy perspective, this interaction can happen in two ways: elastically or inelastically. When an incident particle/electron is expelled by the material, inelastically or elastically, it's refered as a backscattered electron. When an incident electron penetrates into the material it can transfer part of its energy to the material's electrons. Such newly excited electron can transit from the material to the vacuum, i.e., be emittted by the material ; these electrons are refered as secondary electrons.

Thus, it is possible to determine the Total Electron Emission Yield (TEEY), which consists of the ratio between the incident electron impacting the material and electrons emitted by the material, including secondary and backscattered electrons [6].

Measuring TEEY on dielectric materials remains problematic to this day, prone to numerous artifacts. Indeed,

dielectrics are susceptible to trapping electrical charges (positive or negative) subsequent to irradiation and electron emission. This charge can influence both the trajectories of emitted and incident electrons in vacuum [7] as well as the transport of secondary electrons within the material [8]. To limit this charge to an acceptable threshold, such that it does not alter the measurements, it is often necessary to restrict the flux of incident electrons using a pulsed electron beam. Nevertheless, despite these precautions, charging effects are observed depending on the dielectric's thickness (capacitance) and resistivity. A measurement method has been proposed to evaluate this charge during TEEY measurements and to ensure that it does not reach critical values, based on the use of a Kelvin probe potential sensor [9]. More recently, Mata et al [10], revisited and improved this method, demonstrating its interest for TEEY measurements on dielectric materials. However, a major obstacle to the generalization of this method lies in the installation of a Kelvin Probe (KP) within an ultra-high vacuum chamber, mostly due to geometric constraints for existing facilities. We propose here a significant improvement of this method thanks to three key aspects:

(i) The potential probe (KP) is no longer installed in the vacuum chamber [10, 11] but rather in ambient atmosphere (outside the vaccum chamber).

(ii) Charge evaluation is performed during electron irradiation hence in sync with the TEEY measurement, without requiring displacement of either the sample or its holder. This reduces risks of experimental artifact due to charge modification during sample/holder movement.

(iii) Calculation and measurement of the capacitance of the sample-holder-analysis chamber system are no longer necessary, reducing again a possible uncertainty source.

In this paper we first describe the proposed experimental setup, highlighting its adventages compared to existing insitu KP method. We confirm its validity on well-established conductive materials with measurements on a copper sample comparing with the most commonly used method, known as the sample current method [9,10]. Finally, we discuss TEEY measurements on a dielectric material - DuPont™ Kapton® B- using our new setup and we compare our results with literature data.

## 2. Material and Methods

### *2.1 Conventional Methods*

To highlight the advantages of our novel approach, we first provide a brief overview of existing methods for measuring the TEEY.

Traditional TEEY measurement techniques rely on the principle of charge or current conservation. An electron beam with intensity $I_0$ is directed at the sample surface, causing electron emission and generating a current $I_E$. The resulting excess or deficit of charge either flows to the ground as a sample current ($I_S$) in conductive materials or remains trapped in insulating materials. In both cases, a current ($I_S$) circulates between the sample holder and ground.

The physical interpretation of $I_S$ differs between conductive and insulating samples:

- For conductors, $I_S$ represents the discharge of excess injected charge.
- For insulators, $I_S$ corresponds to a displacement or induced current.

TEEY is typically determined by measuring two out of the three currents:

- $I_0$, using a Faraday cup or a positively polarized sample.
- $I_E$, with a positively biased collector surrounding the sample.
- $I_S$, the current flowing to the sample holder.

This method is straightforward and well-suited for conductive materials. However, it presents significant limitations for non-conductive materials, such as dielectrics electrical floating conductors. Charge trapping during irradiation can distort measurements, particularly in thicker dielectrics, where the lower capacitance between the sample surface and holder leads to a larger surface potential for a given trapped charge and hence higher electric field into the vacuum.

To mitigate charging effects, short electron pulses are often used, but this does not entirely eliminate the issue. It is crucial to assess the charge amplitude during measurement and ensure its neutralization. The shape and stability of the collector or sample current signal can provide qualitative insights into the extent of charge accumulation.

Another limitation of determining TEEY from $I_S$ in thick dielectric samples (on the millimeter scale) has been recently highlighted [10]. The electrostatic influence between trapped charges and their image charges on the sample holder is not total, leading to an underestimation of TEEY.

To overcome this issue, a method based on direct measurement of trapped charge has been developed [9]. This approach involves measuring the surface potential variation ($\Delta V_S$) induced by the trapped charge ($Q_t$) with the help of an in-situ Kelvin probe:

$$Q_t = C.\Delta V_s \qquad (1)$$

Given the incident charge $Q_0$, TEEY is then derived as:

$$TEEY = 1 - \frac{Q_t}{Q_0} = 1 - \frac{C.\Delta V_S}{Q_0} \qquad (2)$$

Where C is the electrical capacitance of the system, either calculated for simple geometries [11] or measured in-situ [10, 11].

Despite its advantages, this method has two major drawbacks:

- Integration complexity: Incorporating the Kelvin probe into the vacuum chamber requires modifications that may be incompatible with spatial constraints.
- Time-consuming measurement process: Irradiation and $V_S$ measurement occur in successive steps, requiring tedious and time-intensive repositioning of the KP or sample holder.

## *2.2 Proposed Method*

### *Experimental arrangement*

To overcome these limitations, we propose an experimental setup illustrated in Figure 1.

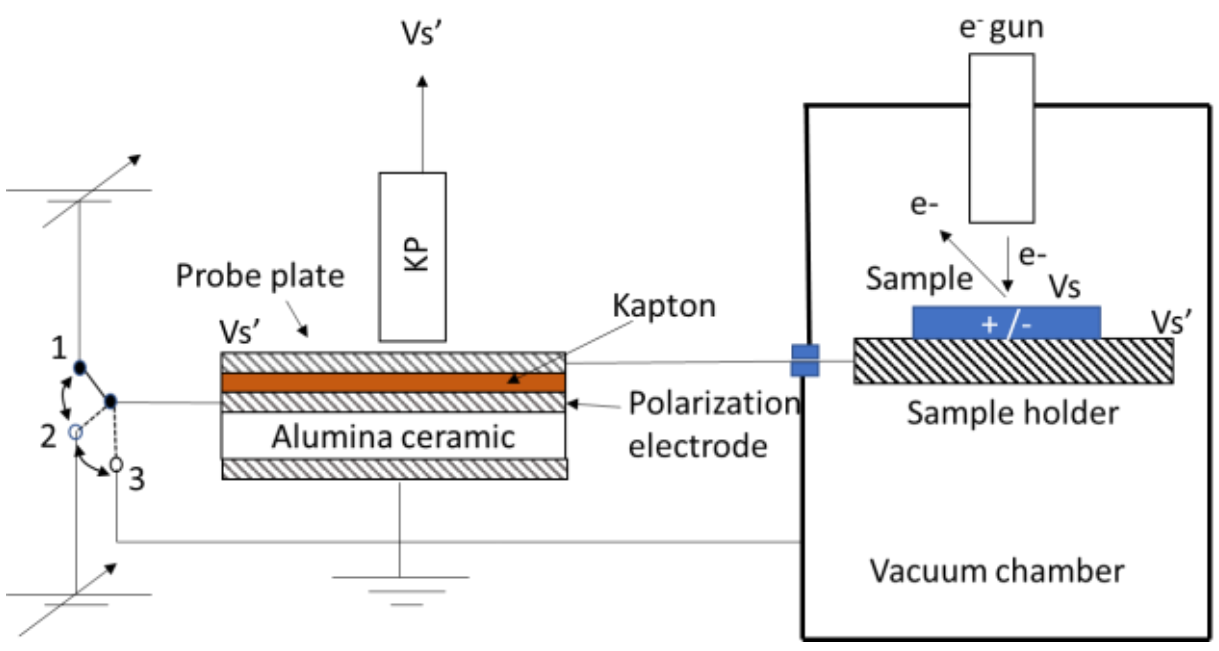

Figure 1 : experimental setup

The metallic or dielectric sample is placed on a metallic sample holder, which is electrically connected to a surface potential measurement device located outside the vacuum chamber. The measurement device consists of a first metallic measurement plate connected to the sample holder (probe plate). This plate is electrically insulated by a Kapton film from a second metallic plate, which we refer to as the "polarization electrode." This polarization electrode is insulated from the common ground of the vacuum chamber by an alumina plate.

In this system, the probe plate is thus electrically floating and is at the same potential of the sample holder, which is also floating. The polarization electrode can be connected, as needed, to either a negative bias (1), a positive bias (2), or ground (3). A Treck 323 Kelvin probe, connected to Trek Model 323 electrostatic voltmeter faces the probe plate. The vacuum chamber is equipped with Kimball Physics EGL2 2022 electron gun.

### *Methodology*

When the surface of the sample is irradiated by a pulse of incident electrons with charge $Q_0$, secondary and backscattered electrons, representing a charge $Q_E$, are emitted into the vacuum. The difference between the emitted charge and the incident charge remains trapped; this is the charge $Q_t$.

This charge induces a potential $V_S$ at the surface of the sample through capacitive coupling. The same charge also produces a surface potential $V_S'$ on the probe plate, where $V_S'$ is proportional to $V_S$. $V_S$ itself is proportional to $Q_t$ (Equation 1).

Thus, there is a linear relationship between $Q_t$ and $V_S'$:

$$Q_t = C'.V_S' \qquad (3)$$

Where C′ is the global capacitance of the system composed of the sample holder, the analysis chamber, and the measurement device.

According to Equation 2:

$$TEEY = 1 - \frac{Q_t}{Q_0}$$

To measure the amount of incident charge, it is common practice to positively polarize the sample holder. Conversely, to measure the amount of emitted charge, the sample holder is usually negatively polarized. In our standard procedure [9,11], we demonstrated that a positive polarization of +27 V was sufficient to measure the incident current and that a negative polarization of -9 V ensured that all electrons emitted by the surface were indeed escaping it [9,11].

In our new setup, the polarization electrode provides these polarizations to the sample holder through the probe plate. Figure 2 shows the variation of the surface potential $\Delta V_S'$ measured by the KP probe on the probe plate as a function of the applied polarization on the polarization electrode ($V_{bias}$).

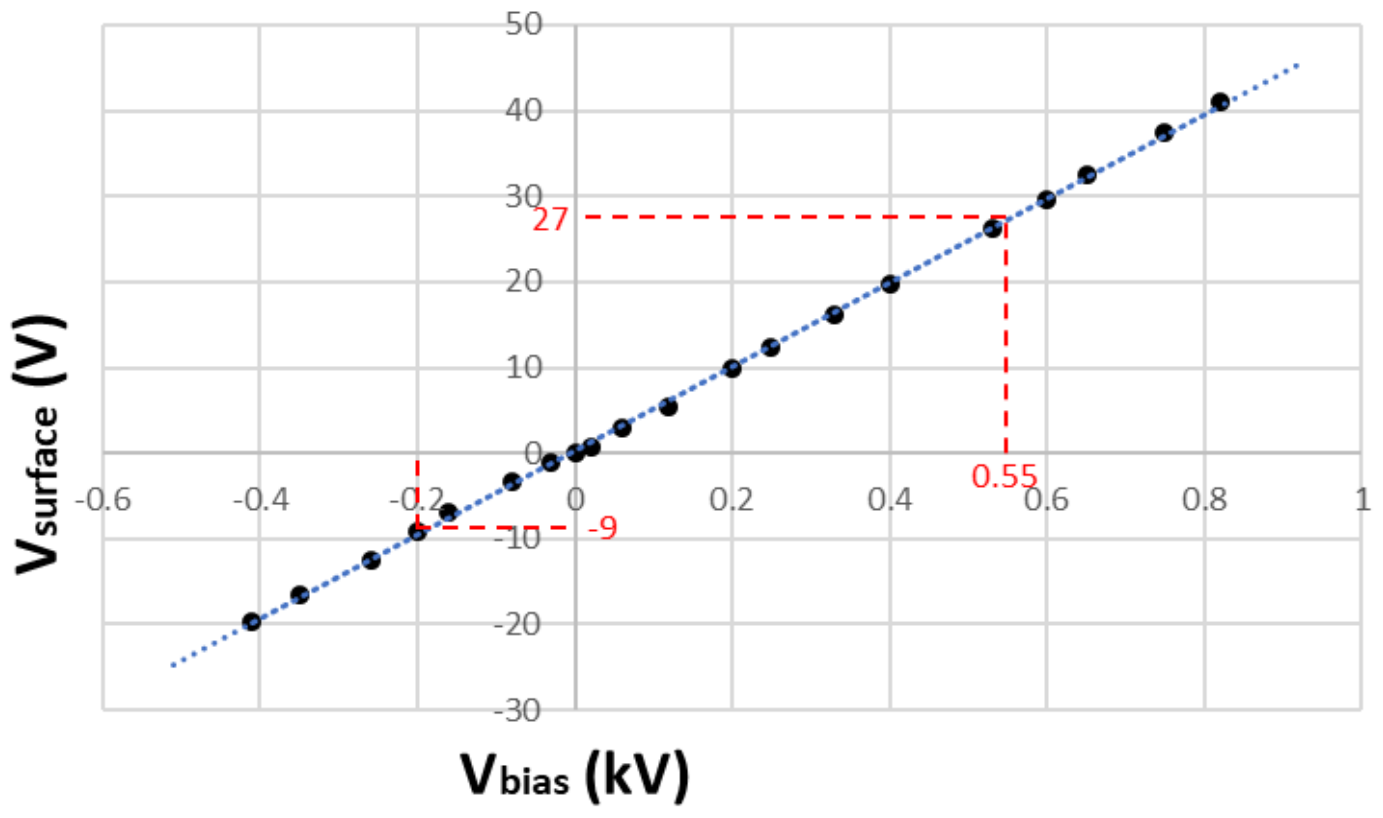

Figure 2 : surface potential on the probe plate as function of the applied potential ($V_{bias}$) on the polarization electrode. Dashed line: linear fit of the experimental data (dots)

Biases of +550V and -200V are applied to the polarization electrode to achieve initial $V_S'$ values of +27 V and -9 V, respectively.

After irradiation by an electron pulse with a bias voltage $V_{bias}$ of +550V, the measured potential variation $\Delta V_S'(0)$ corresponds to the amount of incident charge:

$$Q_0 = C'.\Delta V_S'(0) \qquad (4)$$

Similarly, using the exact same pulse (same amount of incident charge) but applying a polarization of -200V, the measured potential variation $\Delta V_S'(t)$ corresponds to the amount of charge that remains trapped (not emitted), $Q_t$. That is:

$$Q_t = C'.\Delta V_S'(t) \qquad (5)$$

By applying Equation 1, it is therefore possible to determine the TEEY from the sole measurement of potential variations:

$$TEEY = 1 - \frac{\Delta V_S'(t)}{\Delta V_0'(0)} \qquad (6)$$

Another advantage is added, here to the two listed above: it is no longer necessary to calculate or measure the system capacitance, unlike previous protocols [10, 11], reducing a possible uncertainty source.

## 3. Results

### *3.1 Validation on metallic floating sample*

To validate our new method, we first measured the total electron emission yield (TEEY) of a copper sample (99.99 pure Cu obtained from Goodfellow®) that had been exposed to the atmosphere for several months. Metals offer the advantage of accumulating charge when left electrically floating, enabling TEEY measurement using the proposed method. However, when grounded, their TEEY can be determined using the well-established current-based method.

For each incidence energy, the variations in surface potential induced by an electron with 10 pulses of 6ms are measured for both a positive electrode polarization ($\Delta V_S'(0)$) and a negative polarization ($\Delta V_S'(t)$). After each pulse, the surface is discharged by connecting the probe plate to the ground (switch in position 3 in Figure 1). The variations in surface potential are shown in Figure 3. Notably, the surface potential variation induced by charge accumulation consistently remains below 5 V. This ensures that the surface potential of the sample, initially biased at –9 V, remains negative throughout the measurement, thereby preventing any suppression of secondary electron emission.

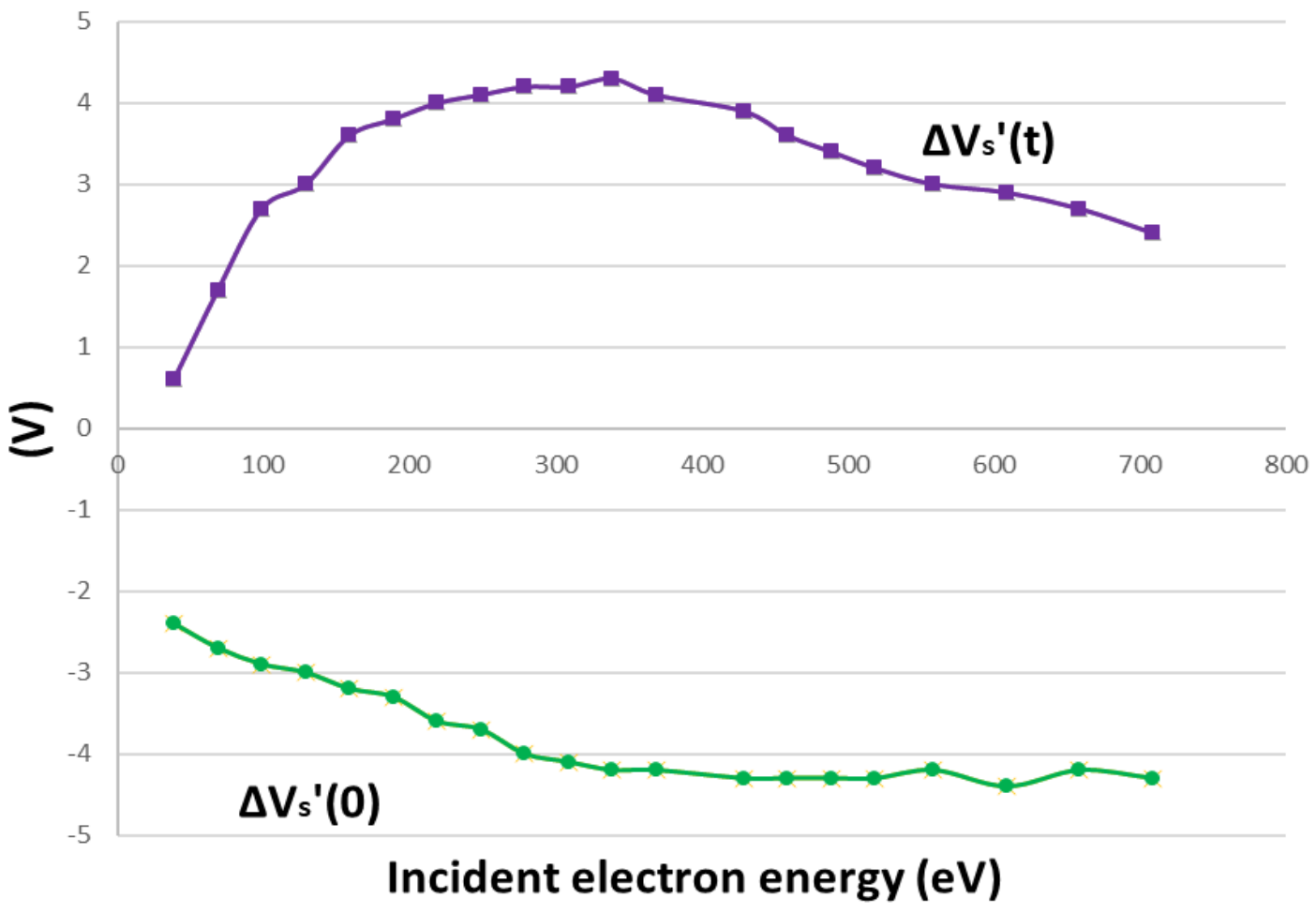

Figure 3: Surface potential variations of the probe plate as function of the incident electron energy

The TEEY curve derived from these measurements using equation 6 is compared to the TEEY measured on the same sample, under the same configuration and within the same chamber, using the conventional sample current method in Figure 4.

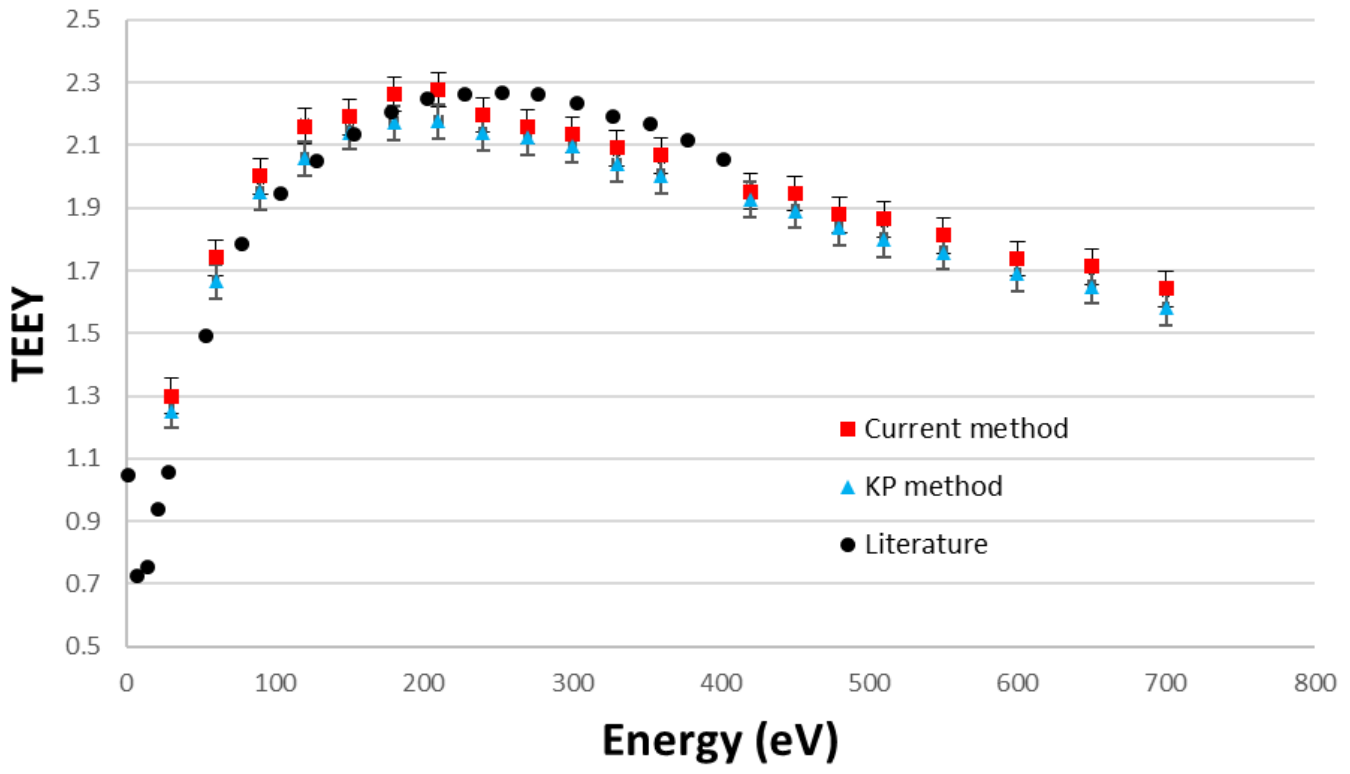

Figure 4: TEEY of Cu (exposed to ambient atmosphere) measured by the proposed new approach (blue triangles), the conventional sample current method (red square) and the curve from Larciprete et al [15] (black circle).

The curves show similar behaviour, with the current method presenting slightly values superiors in comparison to KP method. However, considering the calculated standard deviation (around 2%) for both methods, the error bars overlap. We can conclude that the results obtained on this same copper sample are consistent. The TEEY of Cu exposed to ambient atmosphere measured by Larciprete et al [15] is also presented for comparison. The small difference may be explained by the fact that the surface composition may depends on the storage conditions [16].

### *3.2 Measurement on dielectric sample*

The ex-situ KP method was subsequently applied to a dielectric: DuPont™ Kapton® B polymer (50μm thick). During the measurement of the TEEY using electron with pulses of 6ms the sample surface was discharged using both electrons and a Krypton UV lamp (110 nm), for both negative and positive charging. The effectiveness of the discharging is monitored. One of the main advantages of employing the ex-situ Kelvin Probe (KP) method lies in the continuous monitoring of the surface potential during irradiation, whether by UV light or electrons. This approach not only facilitates the discharge protocol but, more importantly, ensures the effectiveness and reliability of the discharge process.

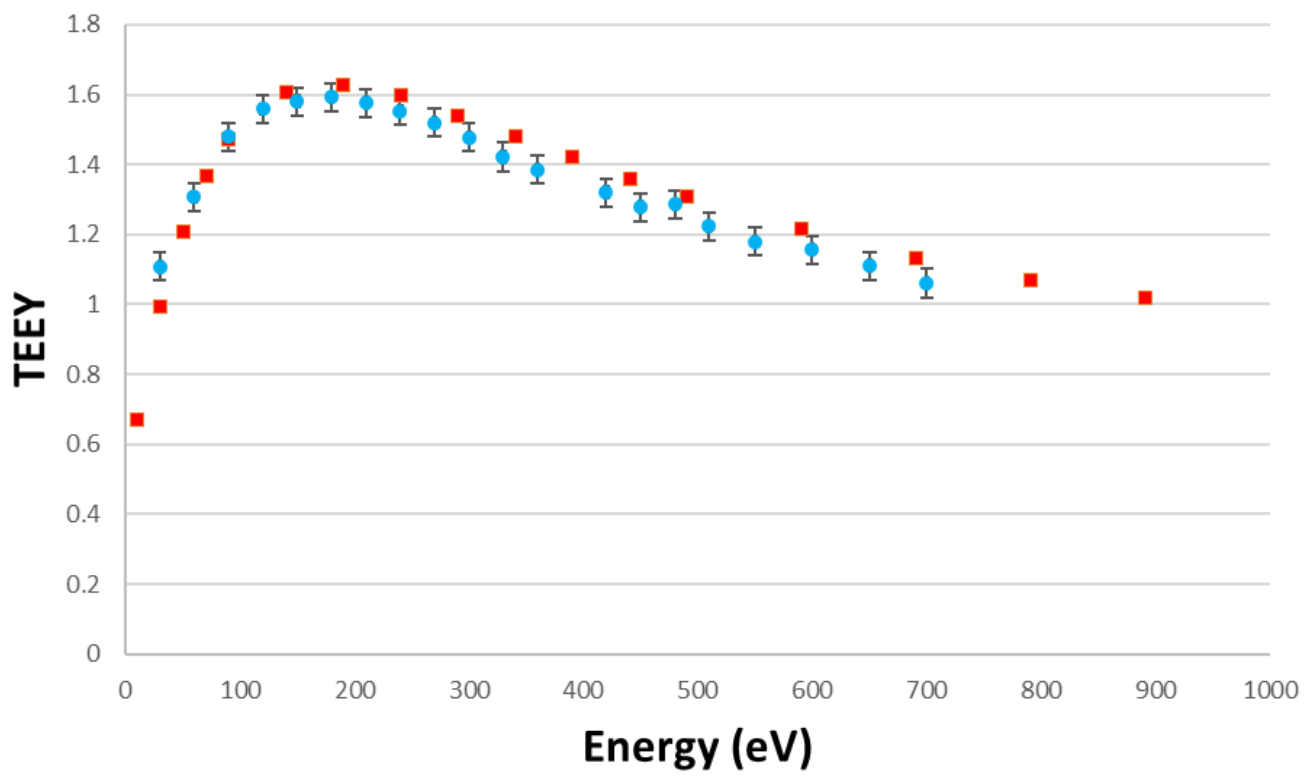

Figure 5: TEEY of Black Kapton measured by the proposed approach (blue circle) and current method (red square).

Figure 5 presents a comparison between our newt setup and a well-established TEEY facility ALCHIMIE [17]. The consistency between the two datasets demonstrates a high level of agreement, thereby reinforcing the reliability and reproducibility of the KP method for this type of material characterization.

## 4.Conclusion

In conclusion, the ex-situ Kelvin Probe (KP) method has proven to be a robust and reliable technique for measuring the total electron emission yield (TEEY) in both conductive and dielectric materials. Beyond its practicality, the method is particularly advantageous for experimental setups where the integration of an internal KP probe is not feasible due to spatial constraints.

A key benefit of this ex-situ approach lies in its simplified formulation, which does not require prior knowledge or calculation of the system's capacitance. This modification reduces potential sources of uncertainty commonly associated with the traditional in-situ KP method, thereby improving the accuracy and reliability of the measurements.

Furthermore, the external configuration, involving a separate probe and polarizing electrode, enhances the clarity and interpretability of the measurement process especially in the case of dielectric samples. One of the main advantages of the KP method compared to current-based techniques is its ability to monitor surface charge concentration in real time through the potential probe. This enables the sample to be discharged between measurements, ensuring consistent and repeatable results with minimal uncertainty.

The principle of the method can be readily extended to the investigation of the emission of various types of charged particles, such as ions, under a wide range of excitation sources, including photons, ions, and others.

Finally, the method offers ease of implementation. While a Kelvin probe was employed in this study, equivalent results can be achieved using a standard multimeter as in [18] for charge monitoring or any comparable voltage measurement device.

## Acknowledgements

This work was supported by the joint research program between CNES and ONERA: PIC Multipactor.